\documentclass[10pt,conference]{IEEEtran}
\IEEEoverridecommandlockouts
\usepackage{cite}
\usepackage{amsmath,amssymb,amsfonts}
\usepackage{algorithm}
\usepackage{algpseudocode}
\usepackage{graphicx}
\usepackage{textcomp}
\usepackage{xcolor}
\usepackage{cite}
\usepackage{comment}
\usepackage{caption}
\usepackage{subcaption}
\usepackage{pgfplots}
\usepackage{tikz}
\usepackage[nolist]{acronym}
\usepackage{siunitx}
\DeclareSIUnit{\belisotropic}{Bi}
\DeclareSIUnit{\dBi}{\deci\belisotropic}
\def\BibTeX{{\rm B\kern-.05em{\sc i\kern-.025em b}\kern-.08em
    T\kern-.1667em\lower.7ex\hbox{E}\kern-.125emX}}
\pgfplotsset{compat=1.18}

\begin{document}
\begin{acronym}
    \acro{faa}[FAA]{Federal Aviation Administration}
    \acro{euro}[EUROCONTROL]{European Organization for the Safety of Air Navigation}
    \acro{icao}[ICAO]{International Civil Aviation Organisation}
    \acro{fci}[FCI]{future communications infrastructure}
    \acro{sesar}[SESAR]{Single European Sky ATM Research}
    
    \acro{atm}[ATM]{air traffic management}
    \acro{dme}[DME]{distance measuring equipment}
    \acro{ldacs}[LDACS]{L-band Digital Aeronautical Communications System}
    \acro{vhf}[VHF]{very high-frequency}

    \acro{ac}[AC]{aircraft}
    \acro{gs}[GS]{ground station}
    \acro{ag}[AG]{air-ground}
    \acro{rl}[RL]{reverse link}
    \acro{fl}[FL]{forward link}
    \acro{tx}[Tx]{transmitter}
    \acro{rx}[Rx]{receiver}
    \acro{uav}[UAVs]{unmanned aerial vehicles}
    
    \acro{msl}[MSL]{mean sea level}
    \acro{nm}[NM]{nautical miles}
    \acro{rf}[RF]{radio frequency}
    
    \acro{tl}[TL]{takeoff \& landing}
    \acro{cd}[CD]{climb \& descent}
    \acro{ec}[EC]{enroute cruise}
    \acro{aoa}[AoA]{angle of arrival}
    
    \acro{fspl}[FSPL]{free-space path loss}
    \acro{mpc}[MPC]{multipath component}
    \acro{los}[LOS]{line-of sight}
    \acro{gmp}[GMP]{ground multipath component}
    \acro{sr}[SR]{specular reflection}
    \acro{lmp}[LMPs]{lateral multipath components}
    
    \acro{upa}[UPA]{uniform planar rectangular array}
    \acro{uca}[UCA]{uniform circular array}
    \acro{sa}[SA]{single-omnidirectional antenna}
    \acro{ma}[MA]{multiple antenna}
    \acro{hpbw}[HPBW]{half-power beamwidth}
    \acro{siso}[SISO]{single input single output}
    \acro{simo}[SIMO]{single input multiple output}
    \acro{miso}[MISO]{multiple input single output}
    \acro{mimo}[MIMO]{multiple input multiple output}
    \acro{su-simo}[SU-SIMO]{single-user single input multiple output}
    \acro{mu-mimo}[MU-MIMO]{multiuser multiple input multiple output}
    \acro{mu-simo mac}[MU-SIMO MAC]{multiuser single input multiple output multiple-access channel}
    \acro{mac}[MAC]{multiple-access channel}
    \acro{noma}[NOMA]{non-orthogonal multiple access}
    \acro{oma}[OMA]{orthogonal multiple access}
    \acro{cgtr}[CGTR]{channel gain and transmission rate}
    \acro{cfo}[CFO]{carrier frequency offset}
    
    \acro{ofdm}[OFDM]{orthogonal frequency-division multiplexing}
    \acro{zf}[ZF]{zero forcing}
    \acro{ssa}[SSA]{single successive algorithm}
    \acro{gsa}[GSA]{group successive algorithm}
    \acro{lgsa}[LGSA]{limited group successive algorithm}
    \acro{vblast}[V-BLAST]{vertical-bell laboratories layered space-time}
    \acro{sic}[SIC]{successive interference cancellation}
    \acro{isu}[ISU]{independent single-user}
    \acro{mmse}[MMSE]{minimum mean squared error}
    \acro{mmse-sic}[MMSE-SIC]{minimum mean squared error with successive interference cancellation}
    \acro{vms}[VMS]{V-BLAST MMSE-SIC}
    \acro{zf-sic}[ZF-SIC]{zero forcing detector with successive interference cancellation}
    \acro{bps}[bps/Hz]{bits per second per hertz}
    
    \acro{snr}[SNR]{signal-to-noise ratio}
    \acro{sir}[SIR]{signal-to-interference-ratio}
    \acro{sinr}[SINR]{signal-to-noise-interference-ratio}
    \acro{ber}[BER]{bit error rate}
    \acro{awgn}[AWGN]{additive white Gaussian noise}
    \acro{qpsk}[QPSK]{quadrature phase shift keying}
    \acro{cr}[CR]{coding rate}
    \acro{rs}[RS]{receiver sensitivity}

    \acro{pdf}[PDF]{probability density function}
    \acro{cdf}[CDF]{cumulative density function}
    
    \acro{doa}[DOA]{direction of arrival}
    \acro{dod}[DOD]{direction of departure}
    
    \acro{cir}[UCA]{uniform circular array}
    \acro{rec}[UPRA]{uniform planar rectangular array}
    \acro{sun}[UFSA]{uniform Fermat spiral array}
  
    \acro{tdma}[TDMA]{time division multiple access}
    \acro{ofdma}[OFDMA]{orthogonal frequency-division multiple access}
    
    \acro{rhs}[RHS]{right hand side}
    
    \acro{td}[TD]{time-division}
    \acro{zc}[ZC]{Zadoff-Chu}
\end{acronym}

\newcommand{\ag}[1]{\textcolor{blue}{#1}}
\newcommand{\ed}[1]{\textcolor{orange}{#1}}
\newcommand{\red}[1]{\textcolor{red}{#1}}
\newcommand{\gre}[1]{\textcolor{green}{#1}}


\setlength{\abovecaptionskip}{2pt}
\setlength{\belowcaptionskip}{2pt}
\setlength{\intextsep}{1mm}
\setlength{\textfloatsep}{1mm}
\setlength{\dblfloatsep}{2mm}
\setlength{\dbltextfloatsep}{2mm}

\newtheorem{theorem}{Theorem}[section]
\newtheorem{corollary}{Corollary}[theorem]
\newtheorem{lemma}[theorem]{Lemma}

\algnewcommand{\algorithmicforeach}{\textbf{for each}}
\algdef{SE}[FOR]{ForEach}{EndForEach}[1]
  {\algorithmicforeach\ #1\ \algorithmicdo}
  {\algorithmicend\ \algorithmicforeach}
\algnewcommand{\LineComment}[1]{\State \(\triangleright\) #1}

\algrenewcommand\algorithmicrequire{\textbf{Input:}}
\algrenewcommand\algorithmicensure{\textbf{Output:}}

\newcommand\Algphase[1]{%
\vspace*{-.4\baselineskip}\Statex\hspace*{\dimexpr-\algorithmicindent-2pt\relax}\rule{0.49\textwidth}{0.1pt}%
\Statex\vspace{-.3\baselineskip}\hspace*{-\algorithmicindent}{#1}%
\vspace*{-.7\baselineskip}\Statex\hspace*{\dimexpr-\algorithmicindent-2pt\relax}\rule{0.49\textwidth}{0.1pt}%
}

\newcommand\AlgphaseV[1]{%
\Statex\hspace*{-\algorithmicindent}{#1}%
\vspace*{-.7\baselineskip}\Statex\hspace*{\dimexpr-\algorithmicindent-2pt\relax}\rule{0.49\textwidth}{0.4pt}%
}




\algnewcommand\algorithmicIfT{\textbf{if}}
\algdef{SE}[IFT]{IfT}{EndIfT}[1]
  {\algorithmicIfT\ #1}
  {\algorithmicend\ \algorithmicIfT}


\algtext*{EndIfT}
\algtext*{EndWhile}
\algtext*{EndForEach}
\algtext*{EndFunction}

\title{Channel Estimation under Large Doppler Shifts and Channel Aging in NOMA-Based  Air-Ground Communications  \\
}



\author{\IEEEauthorblockN{ Ayten G{\"u}rb{\"u}z\IEEEauthorrefmark{1} \IEEEauthorrefmark{2},
Giuseppe Caire\IEEEauthorrefmark{2}~\IEEEmembership{Fellow,~IEEE},
and Michael Walter\IEEEauthorrefmark{1}}
\IEEEauthorblockA{\IEEEauthorrefmark{1}  \textit{Institute of Communications and Navigation, German Aerospace Center (DLR), Wessling, Germany}}
\IEEEauthorblockA{\IEEEauthorrefmark{2}\textit{Faculty of Electrical Engineering and Computer Science, Technical University of Berlin, Berlin, Germany}} 
\IEEEauthorblockA{
Email: ayten.guerbuez@dlr.de, caire@tu-berlin.de, m.walter@dlr.de}
}

\IEEEpubid{\begin{minipage}{\textwidth}\ \\ \\ \\ \\ [14pt]
  \footnotesize © 2026 IEEE. Personal use of this material is permitted.  Permission from IEEE must be obtained for all other uses, in any current or future media, including reprinting/republishing this material for advertising or promotional purposes, creating new collective works, for resale or redistribution to servers or lists, or reuse of any copyrighted component of this work in other works.
\end{minipage}}

\maketitle

\begin{abstract}
The growing number of aircraft, combined with the limited available spectrum, poses a challenge for air traffic management communications. 
One promising solution to the problem of spectrum scarcity is the use of multiple antenna systems with \ac{noma}.
While \ac{noma} techniques enhance spectral efficiency, their application to air traffic management communications is challenged by the high speed of the aircraft (up to \qty{214}{m/s}) and the long communication ranges (up to \qty{250}{\kilo\meter}), resulting in  significant Doppler shifts and low signal-to-noise ratios, respectively. 
This study explores the feasibility of \ac{noma} in air-ground communications by employing a realistic geometry-based stochastic air-ground channel model, derived from dedicated flight measurement campaigns.
We assume multiple aircraft simultaneously transmit data to a ground station. 
Our investigation is twofold. First, we study channel estimation using orthogonal pilot sequences. 
Second, we characterize channel aging, which is defined as the time period after which the previously obtained channel estimation becomes outdated.
The results demonstrate that a Doppler-robust channel estimation method is required to fully exploit the potential of \ac{noma}-based air-ground communications.
Our analysis further reveals that the channel estimate remains valid over extended periods during the climb \& descent  and enroute cruise scenarios. 
\end{abstract}

\section{Introduction}\label{sec:intro}
\acresetall 
The aeronautical communications system between  \ac{ac} pilots and air traffic controllers is essential for exchanging flight-critical information related to \ac{atm}. 
Given that \ac{atm}-supporting communications are classified as ``safety of life", they must operate within protected frequency bands to ensure security. The \ac{icao} recommends using the protected frequency band between \qty{960}{}-\qty{1164}{\mega\Hz} 
in the L-band for future aeronautical communications \cite{icao05,wrc-07}.  A major research framework launched in this direction is the \ac{sesar} \cite{sesar}. 

One of the main challenges is that the protected portion of the L-band spectrum is already populated by other legacy systems. 
Therefore, any new aeronautical communications system must not interfere with existing systems while using the limited spectrum as efficiently as possible to meet the growing demand for air transportation \cite{challenge2}. 
In \cite{ayten_tvt0}, we show that employing \ac{noma} in the reverse link, i.e., from the \ac{ac} to the \ac{gs}, offers a promising solution to spectrum congestion in the protected portion of the L-band.
While the analysis in \cite{ayten_tvt0} is based on the assumption that the \ac{rx} has perfect channel knowledge, this paper evaluates the practical feasibility of achieving this spectral efficiency by focusing on channel estimation and the information outage probability under imperfect channel knowledge at the \ac{rx}.
In \ac{noma} systems, one well-established method for channel estimation  involves multiple users (\ac{ac}) transmitting orthogonal pilot sequences simultaneously \cite{cha_est1, cha_est2, 3GPP_zc}.
However, this approach raises a challenge in the \ac{ag} channel because each transmitted signal undergoes a significant and distinct Doppler shift due to the relative motion of the \ac{ac} with respect to the \ac{gs}. These different Doppler shifts disrupt the orthogonality of the pilot sequences and lead to inter-user interference.
Although each \ac{tx} could pre-compensate for the \ac{cfo} to mitigate this issue, the pre-compensation may be inaccurate, since the \ac{cfo} must be estimated at the \ac{tx} prior to the channel estimation, and may change  by the time channel estimation begins.
Another consideration is that the estimated channel becomes outdated over time due to the mobility of the \ac{ac}, which is referred to as ``channel aging." 
Channel aging can lead to performance degradation and hinder efficient communication.

To the best of our knowledge, channel estimation and imperfect channel knowledge at the \ac{rx} in \ac{noma} systems for \ac{atm} communications have not yet been addressed in the literature. 
Nevertheless, high mobility and long communication ranges are also encountered in other non-terrestrial networks, such as \ac{uav} and satellites~\cite{sat1, sat2, sat3, uav1, uav2, uav3}, where \ac{noma} systems have been studied extensively.
In this context, most studies model channel estimation errors as a stochastic term~\cite{sat1, sat2, sat3, uav3}. Similarly, residual interference in imperfect \ac{sic} is modeled either as a stochastic term~\cite{uav1} or as a fractional coefficient that represents the level of  residual interference~\cite{uav3}.
In contrast to the majority of existing literature, the channel model used in this paper is  more akin to ``ray-tracing" than a conventional stochastic model.
We generate a scattering environment based on the statistical distributions in~\cite{nik_tvt1}, where the parameters are derived from dedicated flight measurement campaigns detailed in~\cite{nik_mea, nik_phd}. 
As the \ac{ac} moves within this scattering environment, the channel evolves. Consequently, the channel estimation error accounts for the Doppler shifts of each individual \ac{ac} and the communication range between the \ac{ac} and the \ac{gs}. Moreover,
we compute the residual noise in the imperfect \ac{sic} detector based on the difference between the channel estimated at the \ac{rx} and the actual channel through which the data symbols propagate. Hence,
the residual interference accounts for the elapsed time since the channel was estimated, as well as, the accuracy of the estimated channel.

%
In this study, we estimate the channel using \ac{zc} sequences  \cite{zc}.
In order to assess the performance of the channel estimator  and the effects of channel aging,  we derive two prominent detectors from the estimated channels \cite{mumimoBook}: 1)~a \ac{zf} detector and 2)~a \ac{mmse-sic} detector.
 We then compute the information outage probability. 
This work has three main contributions.
First, we investigate the impact of \ac{tx} \ac{cfo} pre-compensation accuracy on the channel estimation performance of \ac{zc} sequences, as well as its indirect effect on the detector's performance in the \ac{ag} channel.
Second, we characterize the channel aging for \ac{noma}-based \ac{ag} communications. We analyze how the outage probability of imperfect \ac{sic} changes as the actual channel evolves due to the mobility of the \ac{ac}. 
Furthermore, we identify when the \ac{zf} detector is more preferable than the \ac{mmse-sic} detector in terms of performance.
Lastly, we demonstrate that both the channel estimation performance and the channel aging characteristics of the \ac{noma}-based system vary significantly across the three flight scenarios, \ac{tl}, \ac{cd}, and \ac{ec}, and are strongly influenced by channel propagation characteristics.

Following this introduction, the paper is organized as follows: Section~\ref{sec:system_model} presents the system model. In Section~\ref{sec:background}, we provide background information on the \ac{ag} propagation and introduce the channel model used in our simulations. Moving on to Section~\ref{sec:est}, we explain the channel estimation process, while Section~\ref{sec:outage} outlines how the outage probability is computed for the detectors discussed in this paper. The results are presented in Section~\ref{sec:results}. Finally, Section \ref{sec:conc} summarizes our work and provides an outlook for the future.

\section{System Model} \label{sec:system_model}
 In this study, we consider a \ac{rec} at the \ac{gs} equipped with $M$ antenna elements. The antenna elements are arranged in a $\sqrt{M} \times \sqrt{M}$ grid with half-wavelength spacing.
 We assume that there is a minimum data rate requirement for  reliable transmission of flight-critical messages, denoted as $r_G$.
We suppose that the $K$ \ac{ac}, each with a single antenna, transmit at $r_G$ to the \ac{gs} with equal transmission power. 
To comply with the constraints of the protected L-band spectrum for \ac{atm} communications, the system parameters in this paper follow the \ac{sesar} guidelines \cite{LDACSSpec19}. 
Accordingly, each \ac{ac} transmits at a fixed power of  $P = \qty{41}{dBm}$, and the noise power at each \ac{rx} antenna is assumed to be $N_0 = \qty{-107}{dBm}$. The carrier frequency is $f_c=\qty{987}{\MHz}$, the symbol duration is $T_s=\qty{120}{\micro\second}$ and the longest codeword is $T_C=\qty{3.6}{\milli\second}$.  In line with \cite{LDACSSpec19}, this study considers narrowband, frequency-flat channel propagation.
 In the simulations, we adopt a curved Earth model with a radius of \qty{6371}{km} \cite{Geodetic}. The \ac{gs} is located at the center of a circular cell with a \qty{250}{km} radius and positioned \qty{500}{m} above \ac{msl}, as detailed in \cite{nik_phd}.
We evaluate three flight scenarios: \ac{tl}, \ac{cd}, and \ac{ec}, with their corresponding parameters listed in Table~\ref{tab:scenarios}. 
Within the cell, each \ac{ac} is assigned a random position and a direction of movement, with speeds given in Table~\ref{tab:scenarios}. 
To ensure realistic \ac{ac} spacing and prevent overlaps, a minimum separation of \qty{10}{km} is enforced between any two \ac{ac} in the \ac{cd} and \ac{ec} scenarios, and \qty{1}{km} in the \ac{tl} scenario. 
To compute the outage probability, we perform repeated simulations of the \ac{ac}’s initial positions and movement directions within these constraints.

\begin{table}[t]
\caption{Flight Scenarios}
\label{tab:scenarios}
\begin{center}
\begin{tabular}{|p{2cm}|p{1.9cm}|p{1.3cm}|p{1.5cm}|} 
 \hline 
 \textbf{Scenario} & \textbf{Takeoff \& Landing (TL)} & \textbf{Climb \& Descent (CD)} & \textbf{Enroute Cruise (EC)} \\  
 \hline
 Distance between \ac{ac} and GS & \qty{500}{\meter} - \qty{7.3}{\kilo\meter} & \qty{20}{} - \qty{80}{\kilo\meter} & \qty{80}{} - \qty{250}{\kilo\meter}\\[0.5ex]
 \hline
 Speed &  \qty{88}{m/s} & \qty{171}{m/s} & \qty{214}{m/s} \\ [0.5ex]
 \hline
 Altitude relative to MSL  & \qty{530}{} - \qty{815}{\meter} & \qty{3}{} - \qty{9}{\kilo\meter} & \qty{8}{} - \qty{10.4}{\kilo\meter} \\[0.5ex]
 \hline
\end{tabular}
\end{center}
\end{table}

Due to the mobility of the \ac{ac}, the transmitted signals experience a Doppler shift, resulting in a \ac{cfo}. Following \ac{sesar} guidelines, we assume that each \ac{ac} pre-compensates the \ac{cfo}, denoted as $\Delta f$,  before transmission \cite{LDACSSpec19}. Specifically, the estimated \ac{cfo} at the $k$-th transmitting \ac{ac} is denoted by $\Delta \hat{f}_k$ and calculated as $\Delta \hat{f}_k = \eta \cdot \Delta f_k$, where $\eta \in [-1, 1]$ represents the accuracy of the \ac{cfo} estimation and pre-compensation.
For example, if $\eta=1$ then $\Delta \hat{f}_k=\Delta f_k$, meaning the \ac{cfo} is estimated and pre-compensated perfectly. The details of the \ac{cfo} estimation process are beyond the scope of this paper.

\section{\ac{ag} Propagation and Channel Model}\label{sec:background}

This section summarizes the \ac{ag} propagation characteristics based on findings from the L-band \ac{ag} measurement campaigns \cite{nik_mea, nik_phd} and presents the channel model used in this study, derived from \cite{nik_tvt1}.


\subsection{\ac{ag} Propagation Characteristics}\label{sec:ag_pro}
Based on \cite{nik_mea, nik_phd}, the most dominant \ac{mpc} in the L-band \ac{ag} channel is the \ac{los} path, representing the direct link between the \ac{ac} and the \ac{gs}.
Following this, the \ac{gmp} is typically the second strongest \ac{mpc}. It is a specular reflection, where the reflecting point lies between the \ac{ac} and the \ac{gs}, with its exact position determined by the coordinates of the \ac{ac} and the \ac{gs}. The \ac{gmp} arrives at the \ac{gs} shortly after the \ac{los} signal, with a Doppler frequency that is very similar to the \ac{los} signal. The \ac{gmp} often causes flat fading. Lastly, \ac{mpc}s reflecting off buildings, large structures, or vegetation are called \ac{lmp}. The reflectors that produce \ac{lmp} are typically located near the \ac{gs}. The power of \ac{lmp}  is much lower than that of the \ac{los} and \ac{gmp}.

The analysis in \cite{nik_phd} categorizes the propagation characteristics of the \ac{ag} channel into three main phases of a typical flight scenario: \ac{tl}, \ac{cd}, and \ac{ec}. 
During the \ac{ec} phase, the large distance between \ac{tx} and \ac{rx} results in a similar and  slowly changing \ac{aoa} for the \ac{los} and \ac{gmp} components, causing a very low Doppler spread. In the \ac{cd} phase, the shorter distance between \ac{tx} and \ac{rx} leads to slightly faster \ac{aoa} variations and a marginally higher Doppler spread compared to the \ac{ec} phase. During the \ac{tl} phase, despite lower \ac{ac} speed, the close proximity to the \ac{gs} causes rapid \ac{aoa} changes and greater \ac{aoa} separation between \ac{los} and \ac{gmp}, resulting in the highest Doppler spread.
Overall, considering \ac{mpc}s whose relative power compared to the \ac{los} signal is greater than \qty{-20}{\dB}, the Doppler spread in the \ac{tl} scenario remains below \qty{100}{Hz}, corresponding to a coherence time of approximately $\qty{10}{\milli\second} = 1/\qty{100}{Hz}$ \cite{nik_phd}. Since the \ac{ec} and \ac{cd} scenarios have lower Doppler spreads, their coherence times are longer.
As a result, the channel can be considered approximately constant over the transmission of the longest codeword with a length of \qty{3.6}{\milli\second}, as per \cite{LDACSSpec19}.

\begin{figure}[!t]
     \centering
     \includegraphics[width=0.35\textwidth]{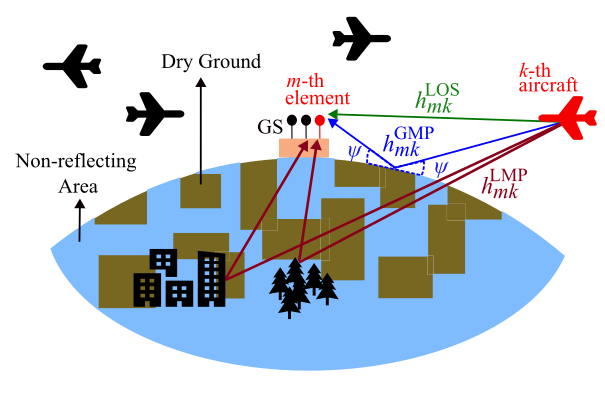}
     \caption{The illustration of the channel model.}
    \label{fig:system}
\end{figure}

\subsection{Channel Model}\label{sec:channel_model}

In this study, we adopt the L-band \ac{ag} channel model proposed in \cite{nik_tvt1}. 
We generate a scattering environment based on the statistical distributions detailed in  \cite{nik_tvt1} and keep it fixed throughout the simulations. Meanwhile, we randomly and repeatedly change the positions of the \ac{ac} within this scattering environment.

In the simulations, we assume that the \ac{los} signal is always present and it is modeled as the shortest possible propagation path between the \ac{ac} and the \ac{gs}. On the other hand, the presence of the \ac{gmp} and \ac{lmp} depends on the position of the \ac{ac} within the scattering environment. We model the \ac{gmp} propagation effect by randomly characterizing the reflecting and non-reflecting areas on the ground, as illustrated in Fig.~\ref{fig:system}. In the simulations, reflecting surfaces cover 50\% of the area around the \ac{gs}, as estimated in \cite{nik_tvt1} for a regional airport. A \ac{gmp} occurs when the ground reflection point lies within a reflecting area. The position of the ground reflection point is determined by the coordinates of both the \ac{rx} and  \ac{tx} antenna. The lateral components that are causing the \ac{lmp} are represented by point reflectors. The number and parameters of the visible \ac{lmp} depend on the \ac{ac} position. The statistical distributions for the \ac{lmp} are derived from information gathered from approximately 130,000 individual reflectors in \cite{nik_tvt1}.

The \ac{mpc}s between the $m$-th antenna element on the \ac{gs} and the $k$-th \ac{ac}, namely the \ac{los}, ${h}^{\text{LOS}}_{mk}$, the \ac{gmp}, ${h}^{\text{GMP}}_{mk}$, and the \ac{lmp}, ${h}^{\text{LMP}}_{mk}$,  are calculated as follows:

\begin{equation}
    {h}^{\text{LOS}}_{mk}[n]=\alpha^{\text{LOS}}_{mk} \exp{ \left(-j2\pi\frac{d^{\text{LOS}}_{mk}[n]}{\lambda}\right)}\exp{ \left(j2\pi \nu_{k}^{\text{LOS}} n \right)} \text{ ,}
\end{equation}
\begin{equation}
    {h}^{\text{GMP}}_{mk}[n]=\rho_{v}\alpha^{\text{GMP}}_{mk}\exp{\left(-j2\pi\frac{d^{\text{GMP}}_{mk}[n]}{\lambda}\right)}\exp{ \left(j2\pi \nu_{k}^{\text{GMP}} n \right)} \text{ ,}
\end{equation}
\begin{align}
    \begin{split}
        {h}^{\text{LMP}}_{mk}[n]=\sum_{l=1}^{L} \hat{\alpha}^{l}_{mk}  \exp{\left(-j2\pi\frac{d^{l}_{mk}[n]}{\lambda}\right)}\exp{ \left(j2\pi \nu_{k}^{l} n \right)} \text{ ,}
    \end{split}
\end{align}
where $n$ is the discrete time index and $L$ is the number of visible lateral reflectors at the \ac{ac}'s given location. The path strengths for the \ac{los} and \ac{gmp} channels, $\alpha_{mk}^{\text{LOS}}$ and $\alpha_{mk}^{\text{GMP}}$, are calculated using the \ac{fspl} formula. Meanwhile, the path strengths of \ac{lmp}, $\hat{\alpha}^{l}_{mk}$, are derived as explained in~\cite{nik_tvt1}. 
The phase of each \ac{mpc} is determined by the wavelength $\lambda$ and its path length $d_{mk}[n]$, which varies with the time index $n$.
The Doppler frequency of each \ac{mpc}, $\nu_{k}$, is determined by the carrier frequency and the time derivative of its corresponding path length.
The vertical reflection coefficient for the \ac{gmp} is denoted by $\rho_{v}$. When the ground reflection point is within a non-reflective area, $\rho_{v}=0$. Otherwise, $\rho_{v}$ is computed using the formula from \cite{Parsons}, accounting for grazing angle and electromagnetic properties of dry ground \cite{ITU}.

As a result, the discrete-time channel at the $n$-th time index between the $m$-th antenna element and  the $k$-th \ac{ac}, ${h}_{mk}[n]$, is calculated by
\begin{equation}
    {h}_{mk}[n]={h}^{\text{LOS}}_{mk}[n]+{h}^{\text{GMP}}_{mk}[n]+{h}^{\text{LMP}}_{mk}[n]\text{.}
\end{equation}
The vector representing the channel between the antenna array on the \ac{gs} and the $k$-th \ac{ac} is given by \mbox{$\textbf{h}_{:k}[n]=[{h}_{1k}[n], {h}_{2k}[n],\dots,{h}_{Mk}[n]]^T\in \mathbb{C}^{M}$}, and the channel matrix between the $K$ \ac{ac} and the \ac{gs} is \mbox{$\textbf{H}_n=[\textbf{h}_{:1}[n], \textbf{h}_{:2}[n], \dots, \textbf{h}_{:K}[n]]\in \mathbb{C}^{M\times K}$}.

\section{Channel Estimation}\label{sec:est}


Within the channel coherence time, the channel matrix can be approximated as $\textbf{H}_n\approx\overline{\textbf{H}}\Lambda_n$, where $\overline{\textbf{H}}\in \mathbb{C}^{M\times K}$ is the slowly varying complex channel gains, and $\Lambda_n \in \mathbb{C}^{K\times K}$  is a diagonal matrix modeling the \ac{cfo}s caused by the Doppler shifts. Specifically, the diagonal elements of $\Lambda_n$ are equal to \mbox{$ [\exp{(j2\pi\Delta f_1 n)},\exp{(j2\pi\Delta f_2 n)}, \dots,\exp{(j2\pi\Delta f_K n)}]$}. We assume that each \ac{ac} pre-compensates for the \ac{cfo}. The diagonal matrix $\hat{\Lambda}\in \mathbb{C}^{K\times K}$ represents the \ac{ac}'s \ac{cfo} pre-compensation with diagonal elements $ [\exp{(-j2\pi\Delta\hat{f}_1 n)},\exp{(-j2\pi\Delta\hat{f}_2 n)}, \dots,\exp{(-j2\pi\Delta\hat{f}_K n)}]$. Accordingly, we denote the uncompensated \ac{cfo} as $\Lambda_n^e = \Lambda_n\hat{\Lambda} \in \mathbb{C}^{K\times K}$. Notice that when the \ac{cfo} is perfectly pre-compensated, $\Lambda_n^e$ becomes the identity matrix, i.e., $\Lambda_n^e=\textbf{I}_K$.

In order to estimate $\overline{\textbf{H}}$, we consider the use of orthogonal \ac{zc} sequences~\cite{zc}.
For simplicity, the system is assumed to have perfect time synchronization.
The set of pilot sequences transmitted by the $K$ \ac{ac} is \mbox{$\sqrt{P}\Phi \in \mathbb{C}^{K\times \tau}$}, where $\tau$ is the sequence length, and the $k$-th row of $\Phi$ corresponds to the sequence transmitted by the \mbox{$k$-th} \ac{ac}. The sequences satisfy the orthogonality condition $\Phi\Phi^{\text{H}}=\textbf{I}_K$, where the superscript H refers to the conjugate transpose. 

The received pilot symbols for \textit{one channel use} of the discrete-time baseband complex channel model are $\textbf{r}_{:b}$ and computed by
\begin{align}\label{eq:zc_trans}
    \begin{split}
       \textbf{r}_{:b}=& \textbf{H}_b\hat{\Lambda}\Phi_{:b}\sqrt{P}+\textbf{z}_{:b} \\
       =& \overline{\textbf{H}}\Lambda_b^e \Phi_{:b}\sqrt{P}+\textbf{z}_{:b} \text{ ,}\quad b=1,\dots,\tau
    \end{split}
\end{align}
where $\Phi_{:b} \in \mathbb{C}^{K}$ is the $b$-th column of $\Phi$, and $\mathbf{z}_{:b} \in \mathbb{C}^{M}$ is complex Gaussian noise $\mathcal{CN}(0, N_0\mathbf{I}_M)$. The channel during the transmission of $\Phi_{:b}$ is approximated as $\mathbf{H}_b \approx \overline{\mathbf{H}}\Lambda_b$. We define $\mathbf{R} = [\mathbf{r}_{:1}, \mathbf{r}_{:2}, \dots, \mathbf{r}_{:\tau}] \in \mathbb{C}^{M\times \tau}$ as the complete received matrix and $\mathbf{Z} = [\mathbf{z}_{:1}, \dots, \mathbf{z}_{:\tau}] \in \mathbb{C}^{M\times \tau}$ as the additive noise matrix.

Notice in \eqref{eq:zc_trans} that when the \ac{cfo} is not perfectly compensated, i.e., $\Lambda_b^e \neq \mathbf{I}_K$, the transmitted pilot symbols are distorted. We denote these modified symbols as \mbox{$\Phi_{:b}^e=\Lambda_b^e \Phi_{:b}\in \mathbb{C}^{K\times 1}$}, which aggregate into the effective pilot matrix $\Phi^e=[\Phi_{:1}^e, \Phi_{:2}^e, \dots,\Phi_{:\tau}^e] \in \mathbb{C}^{K \times \tau}$. In this case, the pilot sequences are de-orthogonalized, such that $\Phi^e\Phi^{\text{H}}\neq\textbf{I}_K$.
When the receiver estimates the channel using the original pilot symbols $\Phi$, the resulting channel estimate is given by
\begin{align}\label{eq:h_est_final}
    \begin{split}
        \hat{\mathbf{H}} &= \mathbf{R}\Phi^{\text{H}}\frac{1}{\sqrt{P}} \\
        &= \overline{\mathbf{H}}\Phi^e\Phi^{\text{H}} + \frac{1}{\sqrt{P}}\mathbf{Z}\Phi^{\text{H}} \text{.}
    \end{split}
\end{align}
According to \eqref{eq:h_est_final}, the sequence length $\tau$ affects the estimation in two opposing ways. A larger $\tau$ effectively averages out the Gaussian noise; however, it also increases sensitivity to uncompensated \ac{cfo}, $\Lambda_b^e$.  Specifically, the off-diagonal elements of $\Phi^e\Phi^{\text{H}}$ increase, which introduces inter-user interference that degrades the channel estimate.
This estimation error limits the reliability of the detectors, which are computed based on the channel estimate to decode the data symbols.

\section{Outage Probability of Detectors}\label{sec:outage}
After the transmission of the pilot symbols, the $K$ \ac{ac} transmit the data symbols. For \textit{one channel use} of the discrete-time baseband complex channel model, the data symbols received at the \ac{gs} are denoted by $\mathbf{y}_\gamma \in \mathbb{C}^{M}$ and are given by
\begin{align}
    \begin{split}
       \textbf{y}_\gamma=&\textbf{H}_\gamma\hat{\Lambda}\textbf{x}_\gamma+\textbf{z}_\gamma \\
       =&\overline{\textbf{H}}_u\Lambda_\gamma^e\textbf{x}_\gamma+\textbf{z}_\gamma\text{ ,}\quad  \gamma=1,\dots,\Gamma
    \end{split}
\end{align}
where $\textbf{x}_\gamma\in \mathbb{C}^{K}$ is the vector of channel inputs transmitted by the $K$ \ac{ac}, and $\Gamma$ is the codeword length. 
As explained in Section~\ref{sec:ag_pro}, the channel coherence time is longer than the duration of the longest codeword.
Accordingly, the channel matrix during the transmission of the $\gamma$-th symbol can be  approximated as $\textbf{H}_\gamma \approx \overline{\textbf{H}}_u\Lambda_\gamma$, where $u$ indexes the codewords. We assume that $\overline{\textbf{H}}_u$ changes from codeword to codeword but remains constant within each codeword.



We consider that the \ac{gs} decodes the received data symbols, $\textbf{y}_\gamma$,  using detectors computed from the estimated channel matrix, $\hat{\textbf{H}}$, which is derived in~\ref{sec:est}. 
In this section, we compute the minimum achievable outage probability, $\mathcal{P}_{\text{out}}$, for the following prominent detectors: \ac{zf} and \ac{mmse-sic}  \cite{mumimoBook}. 
In the case of \ac{sic} schemes, $\mathcal{P}_{\text{out}}$ depends additionally on the decoding order \cite{mimoBook1}.
In \cite{ayten_tvt0}, we proved that the well-known \ac{vblast} algorithm introduced in \cite{vblast} finds the optimal decoding order that minimizes the outage probability, when the users (\ac{ac})  transmit at an equal rate. Based on this result, in this paper we adopt the \ac{vms} detector.
As the channel evolves over time, the previously computed detector becomes outdated, leading to an increase in $\mathcal{P}_{\text{out}}$. Consequently, $\mathcal{P}_{\text{out}}$ is a function of the time elapsed since the last channel estimation.




\subsection{Zero Forcing (ZF) Detector}
\ac{zf} is known for its simplicity and resilience to imperfect channel estimation. However, in low \ac{snr} conditions, it amplifies noise due to instability in matrix inversion in poorly conditioned channels, which leads to significant performance degradation.

The \ac{zf} detector $ \textbf{G}_{\text{ZF}} \in \mathbb{C}^{K\times M}$ is computed by  \mbox{$\textbf{G}_{\text{ZF}} =(\hat{\textbf{H}}^{\text{H}}\hat{\textbf{H}})^{-1}\hat{\textbf{H}}^{\text{H}}$} \cite{mumimoBook}. Accordingly, the symbol transmitted by the $k$-th  \ac{ac} is estimated by 
\begin{equation}
\hat{x}_{k,\gamma} \exp{({j\theta_{k,\gamma}})} = \mathbf{g}_{\text{ZF},k:} \left( \overline{\textbf{H}}_u \Lambda_\gamma^e \mathbf{x}_\gamma + \mathbf{z}_\gamma \right)
\end{equation}
where $\textbf{g}_{\text{ZF},k:}$ is the $k$-th row of $\textbf{G}_{\text{ZF}}$, and $\theta_{k,\gamma}=2\pi \Delta f_k^e\gamma T_s$ represents the phase rotation accumulated up to the $\gamma$-th symbol due to the uncompensated \ac{cfo}, $\Delta f_k^e=\Delta f_k - \Delta \hat{f}_k$.
We assume that this phase rotation is perfectly estimated and neutralized at the \ac{gs} using demodulation reference signals~\cite{5g}. Consequently, the equalized symbols are frequency-synchronized before the hard-decision decoding stage.

The achievable rate of the $k$-th \ac{ac} during the transmission of the $u$-th codeword using the \ac{zf} detector, denoted as $R^{\text{ZF}}_{k}[u]$, is calculated by
\begin{align}\label{eq:zf_out}
    \begin{split}
        R^{\text{ZF}}_{k}[u]=\log_2& \Biggl(1+\dots\\
        &\frac{\lvert\textbf{g}_{\text{ZF},k:}\overline{\textbf{h}}_{:k}[u]\rvert^2P}
    {\sum_{a=1,a\neq k}^{K}\lvert\textbf{g}_{\text{ZF},k:}\overline{\textbf{h}}_{:a}[u]\rvert^2P+ N_0\lvert\lvert \textbf{g}_{\text{ZF},k:}\rvert\rvert^2}\Biggr)\text{,}
    \end{split}
\end{align}
where $\overline{\textbf{h}}_{:k}[u]$ is the $k$-th column of $\overline{\textbf{H}}_u$.
If \mbox{$R_k^{\text{ZF}}[u]<r_G$}, then the \ac{ac} $k$ is in outage. We define the set of users that are in outage at index $u$ as $S_{\text{\text{ZF}}}[u]$, where \mbox{$\forall k \in S_{\text{\text{ZF}}}[u], R_k^{\text{ZF}}[u] < r_G $}. Accordingly, the outage probability for the \ac{zf} detector is \mbox{$\mathcal{P}_{\text{out}}^{\text{ZF}}[u]=\frac{\mathbb{E}[|S_{\text{ZF}}[u]|]}{K}$}, where $\mathbb{E}[\cdot]$ denotes the expectation function. 

\subsection{V-BLAST MMSE-SIC (VMS) Detector}
\ac{sic} detectors can significantly improve performance, particularly in low \ac{snr} conditions. However, when the \ac{rx} has imperfect channel knowledge, errors can accumulate during signal cancellation, making the system susceptible to these imperfections. 

The \ac{vblast} algorithm decodes the  signal with the highest \ac{sinr} at each iteration \cite{vblast}. In \ac{vms}, symbols are decoded using a \ac{mmse} detector at each iteration. The \ac{mmse} filter, \mbox{$\textbf{G}_{\text{MMSE}}\in \mathbb{C}^{K\times M}$}, is computed by \mbox{$\textbf{G}_{\text{MMSE}}=\textbf{Q}\hat{\textbf{H}}^{\text{H}}$}, where \mbox{$\textbf{Q}= \left(\frac{N_0}{P}\textbf{I}_K+\hat{\textbf{H}}^{\text{H}}\hat{\textbf{H}}\right)^{-1}\in \mathbb{C}^{K\times K}$}.
The symbol transmitted by the $k$-th \ac{ac}, $\hat{x}_{k,\gamma}$, can be estimated by \mbox{$\hat{x}_{k,\gamma}\exp{({j\theta_{k,\gamma}})}=\textbf{q}_{k:}\hat{\textbf{H}}^{\text{H}}\textbf{y}_\gamma$},
where $\textbf{q}_{k:}$ is the $k$-th row of $\textbf{Q}$ \cite{mumimoBook}.

\begin{algorithm}
\small
\caption{Outage set with VMS detector}\label{al:vblast}
\begin{algorithmic}[1]
\State $S_{\text{VMS}[u]}=\{1,\dots,K\}$\label{line:def_S}
\For{$o=1,2,...,K$}
\State $\textbf{Q} =\left(\frac{N_0}{P}\textbf{I}_K+\hat{\textbf{H}}^{\text{H}}\hat{\textbf{H}}\right)^{-1}$ \label{line:filter}
\State $i_o=\underset{k}{\text{argmin }}q_{kk}$ \label{line:max}
\State $ \begin{aligned}[t] R_{i_o}^{\text{VMS}}[u]=&\log_{2}\Biggl(1+\dots\\
 &\frac{\lvert\textbf{q}_{i_o:}\hat{\textbf{H}}^{\text{H}}\overline{\textbf{h}}_{:i_o}[u]\rvert^2P}
    {\sum_{ a=1, a\neq o }^{K}\lvert\textbf{q}_{i_o:}\hat{\textbf{H}}^{\text{H}}\overline{\textbf{h}}_{:i_a}[u]\rvert^2P+ N_0\lvert\lvert \textbf{q}_{i_o:}\hat{\textbf{H}}^{\text{H}}\rvert\rvert^2}\Biggr)\end{aligned}$ \label{sic:line:R}
\If{$R_{i_o}^{\text{VMS}}[u] \geq r_G $}
\State Remove $i_o$ from $S_{\text{VMS}}[u]$
\State $\textbf{h}_{:i_o}[u]=\overline{\textbf{h}}_{:i_o}[u]-\hat{\textbf{h}}_{:i_o}$ \label{vblast:line:remove}
\State $\hat{\textbf{h}}_{:i_o}=0$ \label{vblast:line:remove2}
\Else
\State \textbf{break}
\EndIf
\EndFor
\end{algorithmic}
\end{algorithm}

We define $S_{\text{VMS}}$ as the set of \ac{ac} in outage when the \ac{vms} detector is used. Algorithm~\ref{al:vblast} outlines the process of determining $S_{\text{VMS}}[u]$ at codeword index $u$.
In line~\ref{line:def_S}, we first include all transmitting \ac{ac} in the set $S_{\text{VMS}}[u]$. In the first iteration of \textit{for-loop}, we calculate the \ac{mmse} receiver based on the $\hat{\textbf{H}}$ and find the \ac{ac} with the highest \ac{sinr}, which is denoted by $i_o$ (see lines \ref{line:filter} and \ref{line:max}). The achievable rate, $R_{i_o}^{\text{VMS}}[u]$, is calculated in line~\ref{sic:line:R}. If $R_{i_o}^{\text{VMS}}[u]\geq r_G$, then we assume that the $i_o$-th \ac{ac} is successfully decoded, hence $i_o$ is removed from the set $S_{\text{VMS}}[u]$.
In the \ac{vms} decoder, once a symbol is successfully decoded, the decoder subtracts the decoded symbol from the received signal to cancel the interference, i.e., $\textbf{y}_{\gamma}=\textbf{y}_{\gamma} - \hat{\textbf{h}}_{:i_o} x_{i_o,\gamma}\exp{({j\theta_{k,\gamma}})}$.
However, since the \ac{rx} does not have perfect channel knowledge, the residual noise remains at $\textbf{y}_{\gamma}$. We simulate this effect in line~\ref{vblast:line:remove} by subtracting the estimated channel vector, $\hat{\textbf{h}}_{:i_o}$, from the actual channel vector, $\overline{\textbf{h}}_{:i_o}[u]$, under the assumption of perfect \ac{cfo} estimation at the \ac{gs} \ac{rx}.  However, the  \ac{rx} cannot estimate this residual noise, so we must set  the $i_o$-th column of the $\hat{\textbf{H}}$ to zero (see line~\ref{vblast:line:remove2}). 
In subsequent iterations of the \textit{for-loop}, the \ac{mmse} detector uses the updated $\hat{\textbf{H}}$ and $\overline{\textbf{H}}_u$, and the calculation of $R_{i_o}^{\text{VMS}}[u]$ accounts for the remaining noise from the decoded \ac{ac}.
 
If $R_{i_o}^{\text{VMS}}[u] < r_G$, the algorithm stops decoding the remaining \ac{ac}'s in $S_{\text{VMS}}[u]$, and these \ac{ac} are considered in outage. This is because if the \ac{ac} with the highest \ac{sinr} in $S_{\text{VMS}}[u]$ cannot achieve $r_G$, no other \ac{ac} in $S_{\text{VMS}}[u]$ can either, due to mutual interference. Therefore, decoding the remaining \ac{ac}'s in $S_{\text{VMS}}[u]$ becomes impossible.

\section{Numerical Results and Discussion}\label{sec:results}
In this section, we first assess channel estimation under uncompensated \ac{cfo}, and then characterize channel aging in \ac{noma}-based aeronautical communications. 
The simulation parameters are listed in Table~\ref{tab:simulation_parameters}. The parameters for  $P$, $N_0$, $f_c$, $T_s$, and $T_{C}$ follow the \ac{sesar} guidelines \cite{LDACSSpec19}. 

\begin{table}[ht]
    \centering
    \caption{Simulation Parameters}
    \label{tab:simulation_parameters}
    \begin{tabular}{|c|c|c|}
        \hline
        \textbf{Parameter} & \textbf{Symbol} & \textbf{Value} \\ 
        \hline
        Transmit Power          & $P$       & \SI{41}{dBm} \\
        Noise Power Per - \ac{rx} Antenna    & $N_0$     & \SI{-107}{dBm} \\
        Carrier Frequency       & $f_c$     & \SI{987}{MHz} \\
        Symbol Duration         & $T_s$     & \SI{120}{\micro\second} \\
        Codeword Duration       & $T_{C}$       & \SI{3.6}{ms} \\
        Number of \ac{ac}       & $K$       & 4, 8, 16 \\
        Per-\ac{ac} Transmission Rate   & $r_G$     & \SI{2}{bps/Hz} \\
        \ac{gs} \ac{rx} antenna array & --- & \ac{rec} \\
        Number of \ac{rx} Antennas & $M$    & 64 \\
        
        \hline
    \end{tabular}
\end{table}

\begin{figure*}
     \centering
     \begin{subfigure}[b]{0.32\textwidth}
         \centering
         \resizebox{\textwidth}{!}{
%
%
\definecolor{mycolor3}{rgb}{0.85098,0.32549,0.09804}
\definecolor{mycolor2}{rgb}{0.00000,0.44706,0.74118}
\definecolor{mycolor1}{rgb}{1.0, 0.65, 0.0}
\definecolor{mycolor4}{rgb}{0.49412,0.18431,0.55686}
\pgfplotsset{compat=1.10}
\begin{tikzpicture}

\begin{axis}[%
width=4.521in,
height=3.566in,
at={(0.758in,0.481in)},
scale only axis,
xmin=0.8,
xmax=1,
xtick={ 0.80,  0.84, ..., 1.4},
xticklabel style = {font=\LARGE},
xlabel style={font=\color{white!15!black}},
xlabel={\Huge AC Tx CFO Compensation Accuracy ($\eta$)},
ymode=log,
ymin=8e-5,
ymax=1,
ylabel style={font=\color{white!15!black}},
yticklabel style = {font=\LARGE},
ylabel={\Huge $\mathcal{P}_{\text{out}}$ at \qty{3.6}{\milli\second}},
extra y ticks={0.05},
extra y tick style={xticklabel=\pgfmathprintnumber{\tick}},
axis background/.style={fill=white},
xmajorgrids,
ymajorgrids,
yminorgrids,
legend style={at={(0.36,0.42)}, legend cell align=left, align=left, draw=white!15!black}
]

\addplot [color=mycolor1, line width=5.0pt, mark options={solid, mycolor1}] table[row sep=crcr]{%
0.8	0.14065\\
0.81	0.132525\\
0.82	0.124925\\
0.83	0.116525\\
0.84	0.1095\\
0.85	0.10175\\
0.86	0.0943000000000001\\
0.87	0.08725\\
0.88	0.08025\\
0.89	0.073725\\
0.9	0.0678\\
0.91	0.061925\\
0.92	0.0556\\
0.93	0.05\\
0.94	0.04455\\
0.95	0.038675\\
0.96	0.033875\\
0.97	0.02935\\
0.98	0.0261\\
0.99	0.023725\\
1	0.023275\\
};
\addlegendentry{\LARGE VMS $\tau:9$}

\addplot [color=mycolor2, line width=5.0pt, mark options={solid, mycolor1}] table[row sep=crcr]{%
0.8	0.30165\\
0.81	0.279175\\
0.82	0.2589\\
0.83	0.23785\\
0.84	0.218775\\
0.85	0.1999\\
0.86	0.18245\\
0.87	0.1684\\
0.88	0.1532\\
0.89	0.1376\\
0.9	0.12285\\
0.91	0.108275\\
0.92	0.0942499999999999\\
0.93	0.0831499999999999\\
0.94	0.0720499999999999\\
0.95	0.0612\\
0.96	0.051575\\
0.97	0.04125\\
0.98	0.032825\\
0.99	0.02835\\
1	0.0263\\
};
\addlegendentry{\LARGE VMS $\tau:17$}

\addplot [color=mycolor3, line width=5.0pt,  mark options={solid, mycolor3}]
  table[row sep=crcr]{%
0.8	0.576\\
0.81	0.534525\\
0.82	0.490625\\
0.83	0.442075\\
0.84	0.393775\\
0.85	0.349375\\
0.86	0.312375\\
0.87	0.27965\\
0.88	0.2487\\
0.89	0.2217\\
0.9	0.19405\\
0.91	0.1714\\
0.92	0.149025\\
0.93	0.1279\\
0.94	0.1057\\
0.95	0.086225\\
0.96	0.069625\\
0.97	0.05515\\
0.98	0.041175\\
0.99	0.03125\\
1	0.0273\\
};
\addlegendentry{\LARGE VMS $\tau:25$}

\addplot  [color=mycolor1, line width=5.0pt, dashed, mark options={solid, mycolor1}] table[row sep=crcr]{%
0.8	0.02635\\
0.81	0.02585\\
0.82	0.025375\\
0.83	0.02515\\
0.84	0.0249\\
0.85	0.024775\\
0.86	0.024575\\
0.87	0.0241749999999999\\
0.88	0.024275\\
0.89	0.024275\\
0.9	0.0241\\
0.91	0.024125\\
0.92	0.02405\\
0.93	0.0239\\
0.94	0.02385\\
0.95	0.02385\\
0.96	0.023625\\
0.97	0.0236\\
0.98	0.023575\\
0.99	0.023525\\
1	0.02355\\
};
\addlegendentry{\LARGE ZF $\tau:9$}

\addplot  [color=mycolor2, line width=5.0pt, dashed, mark options={solid, mycolor1}] table[row sep=crcr]{%
0.8	0.0397999999999999\\
0.81	0.036225\\
0.82	0.034775\\
0.83	0.033475\\
0.84	0.0322\\
0.85	0.031375\\
0.86	0.03035\\
0.87	0.029675\\
0.88	0.02865\\
0.89	0.028075\\
0.9	0.0276999999999999\\
0.91	0.027475\\
0.92	0.026925\\
0.93	0.02665\\
0.94	0.02635\\
0.95	0.026075\\
0.96	0.02585\\
0.97	0.025825\\
0.98	0.025775\\
0.99	0.025625\\
1	0.025775\\
};
\addlegendentry{\LARGE ZF $\tau:17$}

\addplot [color=mycolor3, line width=5.0pt, dashed,, mark options={solid, mycolor3}]
  table[row sep=crcr]{%
0.8	0.258275\\
0.81	0.206875\\
0.82	0.1623\\
0.83	0.12425\\
0.84	0.087925\\
0.85	0.06005\\
0.86	0.044225\\
0.87	0.037975\\
0.88	0.035775\\
0.89	0.03405\\
0.9	0.03235\\
0.91	0.031175\\
0.92	0.030125\\
0.93	0.0292750000000001\\
0.94	0.0286\\
0.95	0.028125\\
0.96	0.0276\\
0.97	0.02725\\
0.98	0.02695\\
0.99	0.0268\\
1	0.026775\\
};
\addlegendentry{\LARGE ZF $\tau:25$}

\end{axis}
\end{tikzpicture}%
}
         \caption{Takeoff \& Landing (TL)}
         \label{fig:A_TL}
     \end{subfigure}
     \hfill
     \begin{subfigure}[b]{0.32\textwidth}
         \centering
         \resizebox{\textwidth}{!}{
%
%
\definecolor{mycolor3}{rgb}{0.85098,0.32549,0.09804}
\definecolor{mycolor2}{rgb}{0.00000,0.44706,0.74118}
\definecolor{mycolor1}{rgb}{1.0, 0.65, 0.0}
\definecolor{mycolor4}{rgb}{0.49412,0.18431,0.55686}
\pgfplotsset{compat=1.10}
\begin{tikzpicture}

\begin{axis}[%
width=4.521in,
height=3.566in,
at={(0.758in,0.481in)},
scale only axis,
xmin=0.8,
xmax=1,
xtick={ 0.80,  0.84, ..., 1.4},
xticklabel style = {font=\LARGE},
xlabel style={font=\color{white!15!black}},
xlabel={\Huge AC Tx CFO Compensation Accuracy ($\eta$)},
ymode=log,
ymin=8e-5,
ymax=1,
ylabel style={font=\color{white!15!black}},
yticklabel style = {font=\LARGE},
ylabel={\Huge $\mathcal{P}_{\text{out}}$ at \qty{3.6}{\milli\second}},
extra y ticks={0.05},
extra y tick style={xticklabel=\pgfmathprintnumber{\tick}},
axis background/.style={fill=white},
xmajorgrids,
ymajorgrids,
yminorgrids,
legend style={at={(0.36,0.42)}, legend cell align=left, align=left, draw=white!15!black}
]

\addplot [color=mycolor1, line width=5.0pt, mark options={solid, mycolor1}] table[row sep=crcr]{%
0.8	0.219675\\
0.81	0.1947\\
0.82	0.1709\\
0.83	0.151425\\
0.84	0.132725\\
0.85	0.11685\\
0.86	0.101925\\
0.87	0.08945\\
0.88	0.078525\\
0.89	0.0667\\
0.9	0.0569\\
0.91	0.04865\\
0.92	0.04055\\
0.93	0.0322249999999999\\
0.94	0.0238\\
0.95	0.016425\\
0.96	0.010225\\
0.97	0.00522500000000004\\
0.98	0.00229999999999997\\
0.99	0.00102500000000005\\
1	0.000724999999999976\\
};
\addlegendentry{\LARGE VMS $\tau:9$}

\addplot [color=mycolor2, line width=5.0pt, mark options={solid, mycolor1}] table[row sep=crcr]{%
0.8	0.721275\\
0.81	0.692125\\
0.82	0.65735\\
0.83	0.613775\\
0.84	0.5617\\
0.85	0.503775\\
0.86	0.431575\\
0.87	0.354275\\
0.88	0.273525\\
0.89	0.208225\\
0.9	0.16495\\
0.91	0.130475\\
0.92	0.102625\\
0.93	0.079925\\
0.94	0.0622\\
0.95	0.045775\\
0.96	0.03225\\
0.97	0.01845\\
0.98	0.00807500000000005\\
0.99	0.00170000000000003\\
1	0.000600000000000045\\
};
\addlegendentry{\LARGE VMS $\tau:17$}

\addplot [color=mycolor3, line width=5.0pt,  mark options={solid, mycolor3}]
  table[row sep=crcr]{%
0.8	0.860925\\
0.81	0.8461\\
0.82	0.831625\\
0.83	0.815175\\
0.84	0.79445\\
0.85	0.76585\\
0.86	0.733775\\
0.87	0.69295\\
0.88	0.639625\\
0.89	0.56465\\
0.9	0.474225\\
0.91	0.363475\\
0.92	0.2511\\
0.93	0.1731\\
0.94	0.124275\\
0.95	0.087025\\
0.96	0.0588\\
0.97	0.0375\\
0.98	0.0178\\
0.99	0.004525\\
1	0.000399999999999956\\
};
\addlegendentry{\LARGE VMS $\tau:25$}

\addplot  [color=mycolor1, line width=5.0pt, dashed, mark options={solid, mycolor1}] table[row sep=crcr]{%
0.8	0.018775\\
0.81	0.00872499999999998\\
0.82	0.00622500000000004\\
0.83	0.00565000000000004\\
0.84	0.00537500000000002\\
0.85	0.00519999999999998\\
0.86	0.00490000000000002\\
0.87	0.00490000000000002\\
0.88	0.004525\\
0.89	0.00447500000000001\\
0.9	0.00419999999999998\\
0.91	0.00424999999999998\\
0.92	0.00442500000000001\\
0.93	0.00412500000000005\\
0.94	0.00414999999999999\\
0.95	0.00402499999999995\\
0.96	0.00375000000000003\\
0.97	0.00392499999999996\\
0.98	0.00380000000000003\\
0.99	0.00377499999999997\\
1	0.00392499999999996\\
};
\addlegendentry{\LARGE ZF $\tau:9$}

\addplot  [color=mycolor2, line width=5.0pt, dashed, mark options={solid, mycolor1}] table[row sep=crcr]{%
0.8	0.5548\\
0.81	0.5026\\
0.82	0.4421\\
0.83	0.367625\\
0.84	0.2846\\
0.85	0.20865\\
0.86	0.145575\\
0.87	0.091225\\
0.88	0.045525\\
0.89	0.013325\\
0.9	0.00460000000000005\\
0.91	0.00405\\
0.92	0.00360000000000005\\
0.93	0.00342500000000001\\
0.94	0.00322500000000003\\
0.95	0.003\\
0.96	0.00302500000000006\\
0.97	0.00297499999999995\\
0.98	0.00280000000000002\\
0.99	0.00282499999999997\\
1	0.00270000000000004\\
};
\addlegendentry{\LARGE  ZF $\tau:17$}

\addplot [color=mycolor3, line width=5.0pt, dashed,, mark options={solid, mycolor3}]
  table[row sep=crcr]{%
0.8	0.785325\\
0.81	0.768025\\
0.82	0.747925\\
0.83	0.722675\\
0.84	0.68905\\
0.85	0.6396\\
0.86	0.57925\\
0.87	0.5041\\
0.88	0.411775\\
0.89	0.293325\\
0.9	0.182075\\
0.91	0.0986\\
0.92	0.033675\\
0.93	0.00472499999999998\\
0.94	0.00344999999999995\\
0.95	0.00305\\
0.96	0.00290000000000001\\
0.97	0.00260000000000005\\
0.98	0.00267499999999998\\
0.99	0.00260000000000005\\
1	0.00265000000000004\\
};
\addlegendentry{\LARGE  ZF $\tau:25$}

\end{axis}
\end{tikzpicture}%
}
         \caption{Climb \& Descent (CD)}
         \label{fig:A_CD}
     \end{subfigure}
     \hfill
     \begin{subfigure}[b]{0.32\textwidth}
         \centering
         \resizebox{\textwidth}{!}{
%
%
\definecolor{mycolor3}{rgb}{0.85098,0.32549,0.09804}
\definecolor{mycolor2}{rgb}{0.00000,0.44706,0.74118}
\definecolor{mycolor1}{rgb}{1.0, 0.65, 0.0}
\definecolor{mycolor4}{rgb}{0.49412,0.18431,0.55686}
\pgfplotsset{compat=1.10}
\begin{tikzpicture}

\begin{axis}[%
width=4.521in,
height=3.566in,
at={(0.758in,0.481in)},
scale only axis,
xmin=0.8,
xmax=1,
xtick={ 0.80,  0.84, ..., 1.4},
xticklabel style = {font=\LARGE},
xlabel style={font=\color{white!15!black}},
xlabel={\Huge AC Tx CFO Compensation Accuracy ($\eta$)},
ymode=log,
ymin=8e-5,
ymax=1,
ylabel style={font=\color{white!15!black}},
yticklabel style = {font=\LARGE},
ylabel={\Huge $\mathcal{P}_{\text{out}}$ at \qty{3.6}{\milli\second}},
extra y ticks={0.05},
extra y tick style={xticklabel=\pgfmathprintnumber{\tick}},
axis background/.style={fill=white},
xmajorgrids,
ymajorgrids,
yminorgrids,
legend style={at={(0.36,0.42)}, legend cell align=left, align=left, draw=white!15!black}
]

\addplot [color=mycolor1, line width=5.0pt, mark options={solid, mycolor1}] table[row sep=crcr]{%
0.8	0.442425\\
0.81	0.395\\
0.82	0.348275\\
0.83	0.3066\\
0.84	0.268775\\
0.85	0.23965\\
0.86	0.2142\\
0.87	0.189525\\
0.88	0.168775\\
0.89	0.15115\\
0.9	0.132075\\
0.91	0.1152\\
0.92	0.099\\
0.93	0.0845\\
0.94	0.069675\\
0.95	0.0573\\
0.96	0.04655\\
0.97	0.037575\\
0.98	0.031425\\
0.99	0.028025\\
1	0.026775\\
};
\addlegendentry{\LARGE VMS $\tau:9$}

\addplot [color=mycolor2, line width=5.0pt, mark options={solid, mycolor1}] table[row sep=crcr]{%
0.8	0.797725\\
0.81	0.778675\\
0.82	0.75795\\
0.83	0.73875\\
0.84	0.71595\\
0.85	0.682525\\
0.86	0.640425\\
0.87	0.58335\\
0.88	0.514425\\
0.89	0.433425\\
0.9	0.351175\\
0.91	0.276025\\
0.92	0.2202\\
0.93	0.176925\\
0.94	0.142025\\
0.95	0.11095\\
0.96	0.08405\\
0.97	0.05935\\
0.98	0.037925\\
0.99	0.02405\\
1	0.019225\\
};
\addlegendentry{\LARGE VMS $\tau:17$}

\addplot [color=mycolor3, line width=5.0pt,  mark options={solid, mycolor3}]
  table[row sep=crcr]{%
0.8	0.93185\\
0.81	0.915025\\
0.82	0.89585\\
0.83	0.875375\\
0.84	0.853625\\
0.85	0.829175\\
0.86	0.80745\\
0.87	0.7835\\
0.88	0.75345\\
0.89	0.719775\\
0.9	0.669275\\
0.91	0.59415\\
0.92	0.49615\\
0.93	0.374825\\
0.94	0.2643\\
0.95	0.1921\\
0.96	0.135975\\
0.97	0.094525\\
0.98	0.056275\\
0.99	0.0272250000000001\\
1	0.0166500000000001\\
};
\addlegendentry{\LARGE VMS $\tau:25$}

\addplot  [color=mycolor1, line width=5.0pt, dashed, mark options={solid, mycolor1}] table[row sep=crcr]{%
0.8	0.1919\\
0.81	0.1581\\
0.82	0.129025\\
0.83	0.104025\\
0.84	0.088025\\
0.85	0.0785\\
0.86	0.07415\\
0.87	0.07045\\
0.88	0.068875\\
0.89	0.0665249999999999\\
0.9	0.0648\\
0.91	0.0634\\
0.92	0.0629999999999999\\
0.93	0.06185\\
0.94	0.0610000000000001\\
0.95	0.0604749999999999\\
0.96	0.0595250000000001\\
0.97	0.05945\\
0.98	0.05915\\
0.99	0.05875\\
1	0.058575\\
};
\addlegendentry{\LARGE ZF $\tau:9$}

\addplot  [color=mycolor2, line width=5.0pt, dashed, mark options={solid, mycolor1}] table[row sep=crcr]{%
0.8	0.705825\\
0.81	0.677125\\
0.82	0.6439\\
0.83	0.6076\\
0.84	0.5604\\
0.85	0.50065\\
0.86	0.421025\\
0.87	0.330225\\
0.88	0.2434\\
0.89	0.173325\\
0.9	0.11845\\
0.91	0.0777\\
0.92	0.0611\\
0.93	0.056575\\
0.94	0.05425\\
0.95	0.05205\\
0.96	0.050525\\
0.97	0.049575\\
0.98	0.04895\\
0.99	0.04855\\
1	0.047975\\
};
\addlegendentry{\LARGE ZF $\tau:17$}

\addplot [color=mycolor3, line width=5.0pt, dashed,, mark options={solid, mycolor3}]
  table[row sep=crcr]{%
0.8	0.840225\\
0.81	0.82585\\
0.82	0.81015\\
0.83	0.793475\\
0.84	0.77285\\
0.85	0.750675\\
0.86	0.7238\\
0.87	0.6848\\
0.88	0.633325\\
0.89	0.569275\\
0.9	0.4775\\
0.91	0.347775\\
0.92	0.219875\\
0.93	0.12675\\
0.94	0.065075\\
0.95	0.052625\\
0.96	0.0488499999999999\\
0.97	0.0465\\
0.98	0.0449000000000001\\
0.99	0.044175\\
1	0.04425\\
};
\addlegendentry{\LARGE ZF $\tau:9$}

\end{axis}
\end{tikzpicture}%
}
         \caption{Enroute Cruise (EC)}
         \label{fig:A_EC}
     \end{subfigure}
     \caption{ Outage probability, $\mathcal{P}_{\text{out}}$, after \qty{3.6}{\milli\second} (after transmission of one codeword) elapsed since channel estimation for $K=8$ and a \ac{rec} of $M=64$.} 
        \label{fig:accuracy}
\end{figure*}

\begin{figure*}
     \centering
     \begin{subfigure}[b]{0.32\textwidth}
         \centering
         \resizebox{\textwidth}{!}{
%
%
\definecolor{mycolor3}{rgb}{0.85098,0.32549,0.09804}
\definecolor{mycolor2}{rgb}{0.00000,0.44706,0.74118}
\definecolor{mycolor1}{rgb}{1.0, 0.65, 0.0}
\definecolor{mycolor4}{rgb}{0.49412,0.18431,0.55686}
\pgfplotsset{compat=1.10}
\begin{tikzpicture}

\begin{axis}[%
width=4.521in,
height=3.566in,
at={(0.758in,0.481in)},
scale only axis,
xmin=0,
xmax=0.66,
xtick={ 0, 0.1, 0.2,..., 0.6},
xticklabel style = {font=\LARGE},
xlabel style={font=\color{white!15!black}},
xlabel={\Huge Elapsed Time Since Channel Estimation [s]},
ymode=log,
ymin=8e-5,
ymax=0.5,
ylabel style={font=\color{white!15!black}},
yticklabel style = {font=\LARGE},
extra y ticks={0.05},
extra y tick style={xticklabel=\pgfmathprintnumber{\tick}},
ylabel={\Huge $\mathcal{P}_{\text{out}}$},
axis background/.style={fill=white},
xmajorgrids,
ymajorgrids,
yminorgrids,
legend style={at={(0.99,0.43)}, legend cell align=left, align=left, draw=white!15!black}
]

\addplot [color=mycolor1, line width=5.0pt, mark options={solid, mycolor1}] table[row sep=crcr]{%
0.00072	0.00443749999999998\\
0.00144	0.00526249999999995\\
0.00216	0.00586249999999999\\
0.00288	0.00624999999999998\\
0.0036	0.00686249999999999\\
0.006	0.007575\\
0.012	0.00921249999999996\\
0.018	0.0103875\\
0.024	0.011575\\
0.03	0.0125875\\
0.036	0.0135875\\
0.042	0.0144125000000001\\
0.048	0.0154875\\
0.054	0.0155999999999999\\
0.06	0.016325\\
0.12	0.02215\\
0.18	0.0274125\\
0.24	0.0323\\
0.3	0.0368125\\
0.36	0.0412125\\
0.42	0.0464625\\
0.48	0.0505625\\
0.54	0.0544625\\
0.6	0.0588875\\
0.66	0.06265\\
};
\addlegendentry{\LARGE VMS $K:4$}

\addplot [color=mycolor2, line width=5.0pt,  mark options={solid, mycolor2}]
  table[row sep=crcr]{%
0.00072	0.0162125\\
0.00144	0.0193875\\
0.00216	0.0218\\
0.00288	0.0221749999999999\\
0.0036	0.0232625\\
0.006	0.026875\\
0.012	0.0332875\\
0.018	0.0378875\\
0.024	0.040675\\
0.03	0.0430874999999999\\
0.036	0.046425\\
0.042	0.0490375\\
0.048	0.0515625\\
0.054	0.05435\\
0.06	0.0572625\\
0.12	0.0773125\\
0.18	0.0940375\\
0.24	0.1109375\\
0.3	0.1255625\\
0.36	0.1369625\\
0.42	0.1504\\
0.48	0.1618\\
0.54	0.1725625\\
0.6	0.1840875\\
0.66	0.1946125\\
};
\addlegendentry{\LARGE VMS $K:8$}

\addplot [color=mycolor3, line width=5.0pt,  mark options={solid, mycolor3}]
  table[row sep=crcr]{%
0.00072	0.0617124999999999\\
0.00144	0.0708875\\
0.00216	0.07635\\
0.00288	0.08225\\
0.0036	0.085425\\
0.006	0.095825\\
0.012	0.1151625\\
0.018	0.12895\\
0.024	0.1408\\
0.03	0.1487875\\
0.036	0.1572375\\
0.042	0.162375\\
0.048	0.1684125\\
0.054	0.1767125\\
0.06	0.1844375\\
0.12	0.2409375\\
0.18	0.2791875\\
0.24	0.3189375\\
0.3	0.351725\\
0.36	0.377675\\
0.42	0.4052875\\
0.48	0.42835\\
0.54	0.4473\\
0.6	0.467025\\
0.66	0.4865\\
};
\addlegendentry{\LARGE VMS $K:16$}

\addplot  [color=mycolor1, line width=5.0pt, dashed, mark options={solid, mycolor1}] table[row sep=crcr]{%
0.00072	0.00414999999999999\\
0.00144	0.00507500000000005\\
0.00216	0.00586249999999999\\
0.00288	0.00651250000000003\\
0.0036	0.00672499999999998\\
0.006	0.00791249999999999\\
0.012	0.00973749999999995\\
0.018	0.011225\\
0.024	0.0121125\\
0.03	0.012525\\
0.036	0.0130625\\
0.042	0.0134625\\
0.048	0.014\\
0.054	0.0141625\\
0.06	0.014625\\
0.12	0.0183875\\
0.18	0.0223875\\
0.24	0.0260875\\
0.3	0.029475\\
0.36	0.0328375\\
0.42	0.036575\\
0.48	0.039575\\
0.54	0.042425\\
0.6	0.045\\
0.66	0.0477125\\
};
\addlegendentry{\LARGE ZF $K:4$}

\addplot [color=mycolor2, line width=5.0pt,  dashed, mark options={solid, mycolor2}]
  table[row sep=crcr]{%
0.00072	0.0146375\\
0.00144	0.0182\\
0.00216	0.0207125\\
0.00288	0.021725\\
0.0036	0.023275\\
0.006	0.027575\\
0.012	0.0348375\\
0.018	0.0383\\
0.024	0.040675\\
0.03	0.04255\\
0.036	0.0441625\\
0.042	0.044925\\
0.048	0.0461625\\
0.054	0.0477875\\
0.06	0.0485\\
0.12	0.0588125\\
0.18	0.0679875\\
0.24	0.0765625\\
0.3	0.085025\\
0.36	0.093325\\
0.42	0.100475\\
0.48	0.1070125\\
0.54	0.11305\\
0.6	0.119675\\
0.66	0.126475\\
};
\addlegendentry{\LARGE ZF $K:8$}

\addplot [color=mycolor3, line width=5.0pt,  dashed,, mark options={solid, mycolor3}]
  table[row sep=crcr]{%
0.00072	0.0586625\\
0.00144	0.0679125\\
0.00216	0.0745749999999999\\
0.00288	0.0805125\\
0.0036	0.0849124999999999\\
0.006	0.097175\\
0.012	0.115175\\
0.018	0.1248125\\
0.024	0.131075\\
0.03	0.1360625\\
0.036	0.1398\\
0.042	0.1434125\\
0.048	0.1457625\\
0.054	0.149075\\
0.06	0.152575\\
0.12	0.178825\\
0.18	0.1994\\
0.24	0.2193125\\
0.3	0.2357\\
0.36	0.2518625\\
0.42	0.26725\\
0.48	0.2819625\\
0.54	0.2944125\\
0.6	0.307475\\
0.66	0.3187125\\
};
\addlegendentry{\LARGE ZF $K:16$}

\end{axis}
\end{tikzpicture}%
}
         \caption{Takeoff \& Landing (TL)}
         \label{fig:time_TL}
     \end{subfigure}
     \hfill
     \begin{subfigure}[b]{0.32\textwidth}
         \centering
         \resizebox{\textwidth}{!}{
%
%
\definecolor{mycolor3}{rgb}{0.85098,0.32549,0.09804}
\definecolor{mycolor2}{rgb}{0.00000,0.44706,0.74118}
\definecolor{mycolor1}{rgb}{1.0, 0.65, 0.0}
\definecolor{mycolor4}{rgb}{0.49412,0.18431,0.55686}
\pgfplotsset{compat=1.10}
\begin{tikzpicture}

\begin{axis}[%
width=4.521in,
height=3.566in,
at={(0.758in,0.481in)},
scale only axis,
xmin=0,
xmax=0.66,
xtick={0, 0.1, 0.2,..., 0.6},
xticklabel style = {font=\LARGE},
xlabel style={font=\color{white!15!black}},
xlabel={\Huge  Elapsed Time Since Channel Estimation [s]},
ymode=log,
ymin=8e-5,
ymax=0.5,
ylabel style={font=\color{white!15!black}},
yticklabel style = {font=\LARGE},
extra y ticks={0.05},
extra y tick style={xticklabel=\pgfmathprintnumber{\tick}},
ylabel={\Huge $\mathcal{P}_{\text{out}}$},
axis background/.style={fill=white},
xmajorgrids,
ymajorgrids,
yminorgrids,
legend style={at={(0.99,0.43)}, legend cell align=left, align=left, draw=white!15!black}
]

\addplot [color=mycolor1, line width=5.0pt, mark options={solid, mycolor1}] table[row sep=crcr]{%
0.00072	0.000175000000000036\\
0.00144	0.000175000000000036\\
0.00216	0.000175000000000036\\
0.00288	0.000199999999999978\\
0.0036	0.000199999999999978\\
0.006	0.000275000000000025\\
0.012	0.000524999999999998\\
0.018	0.000750000000000028\\
0.024	0.000824999999999965\\
0.03	0.00105\\
0.036	0.0012875\\
0.042	0.00141250000000004\\
0.048	0.00138749999999999\\
0.054	0.00153749999999997\\
0.06	0.00160000000000005\\
0.12	0.00213750000000001\\
0.18	0.00268749999999995\\
0.24	0.00288750000000004\\
0.3	0.00346250000000003\\
0.36	0.00408750000000002\\
0.42	0.00422500000000003\\
0.48	0.00496249999999998\\
0.54	0.00538749999999999\\
0.6	0.00618750000000001\\
0.66	0.00634999999999997\\
};
\addlegendentry{\LARGE VMS $K:4$}

\addplot [color=mycolor2, line width=5.0pt,  mark options={solid, mycolor2}]
  table[row sep=crcr]{%
0.00072	0.000512500000000027\\
0.00144	0.000574999999999992\\
0.00216	0.000574999999999992\\
0.00288	0.000587499999999963\\
0.0036	0.000712500000000005\\
0.006	0.00111249999999996\\
0.012	0.00207500000000005\\
0.018	0.00326249999999995\\
0.024	0.00409999999999999\\
0.03	0.00482499999999997\\
0.036	0.00529999999999997\\
0.042	0.00587499999999996\\
0.048	0.00644999999999996\\
0.054	0.00700000000000001\\
0.06	0.0073375\\
0.12	0.00990000000000002\\
0.18	0.011675\\
0.24	0.0131\\
0.3	0.0141125\\
0.36	0.016025\\
0.42	0.017425\\
0.48	0.019825\\
0.54	0.0208625\\
0.6	0.02275\\
0.66	0.0245375\\
};
\addlegendentry{\LARGE VMS $K:8$}

\addplot [color=mycolor3, line width=5.0pt,  mark options={solid, mycolor3}]
  table[row sep=crcr]{%
0.00072	0.000674999999999981\\
0.00144	0.000950000000000006\\
0.00216	0.00166250000000001\\
0.00288	0.00198750000000003\\
0.0036	0.00236250000000005\\
0.006	0.00444999999999995\\
0.012	0.00856250000000003\\
0.018	0.0129375\\
0.024	0.0164\\
0.03	0.021325\\
0.036	0.0235625\\
0.042	0.0264375\\
0.048	0.0285875\\
0.054	0.0310125\\
0.06	0.03195\\
0.12	0.0409375\\
0.18	0.0492\\
0.24	0.054975\\
0.3	0.0623375\\
0.36	0.0693375000000001\\
0.42	0.074375\\
0.48	0.0808624999999999\\
0.54	0.086125\\
0.6	0.0933125\\
0.66	0.1030875\\
};
\addlegendentry{\LARGE VMS $K:16$}

\addplot  [color=mycolor1, line width=5.0pt, dashed, mark options={solid, mycolor1}] table[row sep=crcr]{%
0.00072	0.000937500000000036\\
0.00144	0.000937500000000036\\
0.00216	0.000962499999999977\\
0.00288	0.00098750000000003\\
0.0036	0.00103750000000002\\
0.006	0.00111249999999996\\
0.012	0.00151250000000003\\
0.018	0.00178750000000005\\
0.024	0.00207500000000005\\
0.03	0.002475\\
0.036	0.00283750000000005\\
0.042	0.00298750000000003\\
0.048	0.00307500000000005\\
0.054	0.00338749999999999\\
0.06	0.00356250000000002\\
0.12	0.00432500000000002\\
0.18	0.00491249999999999\\
0.24	0.00513750000000002\\
0.3	0.00521249999999995\\
0.36	0.00558749999999997\\
0.42	0.00548749999999998\\
0.48	0.00616249999999996\\
0.54	0.00621249999999995\\
0.6	0.00663749999999996\\
0.66	0.00656250000000003\\
};
\addlegendentry{\LARGE ZF $K:4$}

\addplot [color=mycolor2, line width=5.0pt, dashed, mark options={solid, mycolor2}]
  table[row sep=crcr]{%
0.00072	0.00337500000000002\\
0.00144	0.00344999999999995\\
0.00216	0.00356250000000002\\
0.00288	0.00367499999999998\\
0.0036	0.0038125\\
0.006	0.00431250000000005\\
0.012	0.00601249999999998\\
0.018	0.0072875\\
0.024	0.008575\\
0.03	0.00988750000000005\\
0.036	0.0111250000000001\\
0.042	0.0119875\\
0.048	0.012775\\
0.054	0.013625\\
0.06	0.0141250000000001\\
0.12	0.0178\\
0.18	0.0201\\
0.24	0.0209875\\
0.3	0.0213\\
0.36	0.0223375\\
0.42	0.0232125\\
0.48	0.0239\\
0.54	0.0235875\\
0.6	0.024675\\
0.66	0.0248125\\
};
\addlegendentry{\LARGE ZF $K:8$}

\addplot [color=mycolor3, line width=5.0pt, dashed,, mark options={solid, mycolor3}]
  table[row sep=crcr]{%
0.00072	0.0136875\\
0.00144	0.014225\\
0.00216	0.0148125\\
0.00288	0.0154625\\
0.0036	0.0162\\
0.006	0.0187375\\
0.012	0.0241375\\
0.018	0.0301375\\
0.024	0.0348625\\
0.03	0.0403875\\
0.036	0.0441375000000001\\
0.042	0.0478\\
0.048	0.0512625\\
0.054	0.054375\\
0.06	0.0565625\\
0.12	0.0672125\\
0.18	0.07545\\
0.24	0.0773125\\
0.3	0.0809375\\
0.36	0.0840875\\
0.42	0.0859\\
0.48	0.087975\\
0.54	0.090275\\
0.6	0.0930375\\
0.66	0.09565\\
};
\addlegendentry{\LARGE ZF $K:16$}

\end{axis}
\end{tikzpicture}%
}
         \caption{Climb \& Descent (CD)}
         \label{fig:time_CD}
     \end{subfigure}
     \hfill
     \begin{subfigure}[b]{0.32\textwidth}
         \centering
         \resizebox{\textwidth}{!}{
%
%
\definecolor{mycolor3}{rgb}{0.85098,0.32549,0.09804}
\definecolor{mycolor2}{rgb}{0.00000,0.44706,0.74118}
\definecolor{mycolor1}{rgb}{1.0, 0.65, 0.0}
\definecolor{mycolor4}{rgb}{0.49412,0.18431,0.55686}
\pgfplotsset{compat=1.10}
\begin{tikzpicture}

\begin{axis}[%
width=4.521in,
height=3.566in,
at={(0.758in,0.481in)},
scale only axis,
xmin=0,
xmax=0.66,
xtick={ 0, 0.1, 0.2,..., 0.6},
xticklabel style = {font=\LARGE},
xlabel style={font=\color{white!15!black}},
xlabel={\Huge  Elapsed Time Since Channel Estimation [s]},
ymode=log,
ymin=8e-5,
ymax=0.5,
ylabel style={font=\color{white!15!black}},
yticklabel style = {font=\LARGE},
extra y ticks={0.05},
extra y tick style={xticklabel=\pgfmathprintnumber{\tick}},
ylabel={\Huge $\mathcal{P}_{\text{out}}$},
axis background/.style={fill=white},
xmajorgrids,
ymajorgrids,
yminorgrids,
legend style={at={(0.99,0.43)}, legend cell align=left, align=left, draw=white!15!black}
]

\addplot [color=mycolor1, line width=5.0pt, mark options={solid, mycolor1}] table[row sep=crcr]{%
0.00072	0.0210125\\
0.00144	0.0210125\\
0.00216	0.021\\
0.00288	0.020975\\
0.0036	0.0209875\\
0.006	0.0209875\\
0.012	0.0211\\
0.018	0.0212125\\
0.024	0.0212875\\
0.03	0.0214\\
0.036	0.0214625000000001\\
0.042	0.021625\\
0.048	0.0216625\\
0.054	0.021775\\
0.06	0.021925\\
0.12	0.023525\\
0.18	0.0245625\\
0.24	0.0252\\
0.3	0.0255375\\
0.36	0.0261875\\
0.42	0.0264875\\
0.48	0.026525\\
0.54	0.0266999999999999\\
0.6	0.02705\\
0.66	0.026975\\
};
\addlegendentry{\LARGE VMS $K:4$}

\addplot [color=mycolor2, line width=5.0pt,  mark options={solid, mycolor2}]
  table[row sep=crcr]{%
0.00072	0.0266\\
0.00144	0.026525\\
0.00216	0.0265125\\
0.00288	0.0265\\
0.0036	0.0265375\\
0.006	0.0266\\
0.012	0.0267625\\
0.018	0.0269625\\
0.024	0.02725\\
0.03	0.027625\\
0.036	0.0282125\\
0.042	0.028625\\
0.048	0.0292375\\
0.054	0.0297125\\
0.06	0.0303375\\
0.12	0.0349125\\
0.18	0.0373125\\
0.24	0.0382375\\
0.3	0.039175\\
0.36	0.0416875\\
0.42	0.0421\\
0.48	0.0422125\\
0.54	0.0423375\\
0.6	0.0423\\
0.66	0.043825\\
};
\addlegendentry{\LARGE VMS $K:8$}

\addplot [color=mycolor3, line width=5.0pt,  mark options={solid, mycolor3}]
  table[row sep=crcr]{%
0.00072	0.038125\\
0.00144	0.038125\\
0.00216	0.038125\\
0.00288	0.0381625\\
0.0036	0.0381875\\
0.006	0.038575\\
0.012	0.0397625\\
0.018	0.04135\\
0.024	0.0433375\\
0.03	0.04495\\
0.036	0.04625\\
0.042	0.0483749999999999\\
0.048	0.0508125\\
0.054	0.0525\\
0.06	0.0546875\\
0.12	0.0698\\
0.18	0.0785375\\
0.24	0.0840625\\
0.3	0.0881999999999999\\
0.36	0.0916875\\
0.42	0.0955875\\
0.48	0.09555\\
0.54	0.0959375\\
0.6	0.0971375\\
0.66	0.097675\\
};
\addlegendentry{\LARGE VMS $K:16$}

\addplot  [color=mycolor1, line width=5.0pt,  dashed, mark options={solid, mycolor1}] table[row sep=crcr]{%
0.00072	0.0323\\
0.00144	0.0323\\
0.00216	0.0323\\
0.00288	0.0322875\\
0.0036	0.0323\\
0.006	0.0323\\
0.012	0.0323375\\
0.018	0.032525\\
0.024	0.0325875\\
0.03	0.032725\\
0.036	0.032975\\
0.042	0.0330875\\
0.048	0.03315\\
0.054	0.0332875\\
0.06	0.0333875\\
0.12	0.0354875\\
0.18	0.0363125\\
0.24	0.0370125\\
0.3	0.037475\\
0.36	0.0377\\
0.42	0.0383125\\
0.48	0.0383\\
0.54	0.0387875\\
0.6	0.038725\\
0.66	0.03875\\
};
\addlegendentry{\LARGE ZF $K:4$}

\addplot [color=mycolor2, line width=5pt,  dashed, mark options={solid, mycolor2}]
  table[row sep=crcr]{%
0.00072	0.0591625\\
0.00144	0.059175\\
0.00216	0.0591874999999999\\
0.00288	0.0592125\\
0.0036	0.0592\\
0.006	0.0592125\\
0.012	0.0594375\\
0.018	0.0597\\
0.024	0.0599875\\
0.03	0.0606\\
0.036	0.0609875\\
0.042	0.0613875\\
0.048	0.0618625\\
0.054	0.062375\\
0.06	0.0629\\
0.12	0.067425\\
0.18	0.069925\\
0.24	0.0712625\\
0.3	0.0718875\\
0.36	0.0735\\
0.42	0.074475\\
0.48	0.0744875\\
0.54	0.0745749999999999\\
0.6	0.074825\\
0.66	0.0751500000000001\\
};
\addlegendentry{\LARGE ZF $K:8$}

\addplot [color=mycolor3, line width=5.0pt,  dashed,, mark options={solid, mycolor3}]
  table[row sep=crcr]{%
0.00072	0.147825\\
0.00144	0.147825\\
0.00216	0.1478375\\
0.00288	0.14785\\
0.0036	0.1479125\\
0.006	0.1479375\\
0.012	0.148575\\
0.018	0.149575\\
0.024	0.150775\\
0.03	0.1517875\\
0.036	0.1529875\\
0.042	0.1542625\\
0.048	0.15525\\
0.054	0.1564875\\
0.06	0.157875\\
0.12	0.1686\\
0.18	0.17445\\
0.24	0.178225\\
0.3	0.181\\
0.36	0.1839125\\
0.42	0.185875\\
0.48	0.1866\\
0.54	0.18675\\
0.6	0.1869875\\
0.66	0.187925\\
};
\addlegendentry{\LARGE ZF $K:16$}

\end{axis}
\end{tikzpicture}%
}
         \caption{Enroute Cruise (EC)}
         \label{fig:time_EC}
     \end{subfigure}
     \caption{Outage probability, $\mathcal{P}_{\text{out}}$, for different flight scenarios with a \ac{rec} of $M=64$ and varying number of \ac{ac}, $K$. The channel is estimated using \ac{zc} sequences with $\tau=K+1$, assuming perfect \ac{cfo} compensation at the \ac{ac} \ac{tx}, $\eta=1$.} 
        \label{fig:time}
\end{figure*}

\subsection{Channel Estimation Analysis}
In Fig.~\ref{fig:accuracy}, we evaluate the joint impact of the \ac{cfo} compensation accuracy, $\eta$, and pilot sequence length, $\tau$, on channel estimation.
Specifically, the reliability of the channel estimates is evaluated by deriving  detectors based on the channel estimation (see Section~\ref{sec:outage}) and computing the corresponding outage probability, $\mathcal{P}_{\text{out}}$, \qty{3.6}{\milli\second} after channel estimation.
Following the technical specifications in \cite{3GPP_zc},  we choose $\tau$ to be an odd number to ensure a symmetric phase structure in \ac{zc} sequences.
Fig.~\ref{fig:accuracy} demonstrates the trade-off associated with $\tau$, as explained in Section~\ref{sec:est}. A larger $\tau$ improves the quality of channel estimation under low \ac{snr} conditions, as evidenced in the \ac{cd} and \ac{ec} scenarios, where the \ac{snr} is low due to high \ac{fspl}. In these cases, when $\eta=1$, a larger $\tau$ results in a lower $\mathcal{P}_{\text{out}}$. However, as the $\eta$ decreases  sequences with a larger $\tau$ begin to underperform compared to those with a smaller $\tau$. On the other hand, in the \ac{tl} scenario, where the \ac{snr} is already sufficiently high, a larger $\tau$ is not beneficial, even when $\eta=1$. In this case, increasing $\tau$ only increases sensitivity to lower $\eta$ values. Therefore, in the \ac{tl} scenario, a shorter $\tau$ is preferable, since it improves robustness against uncompensated \ac{cfo} and also reduces channel estimation overhead. 

One interesting observation in Fig.~\ref{fig:accuracy} is that, even though the free-space path loss is considerably lower in the \ac{tl} scenario, $\mathcal{P}_{\text{out}}$ is higher than in the \ac{cd} scenario and is comparable to that in the \ac{ec} scenario. We attribute this behavior in the \ac{tl} scenario to the richer scattering environment and the larger variation in the \ac{aoa} of the \ac{mpc}s due to geometry, which leads to increased channel fading. This observation underscores the significant impact of channel propagation characteristics on system performance.

In Fig.~\ref{fig:accuracy}, we observe that as $\eta$ decreases, $\mathcal{P}_{\text{out}}$ increases for both detectors. 
However, we note that this impact is more significant for the \ac{vms} detector than for the \ac{zf} detector across all three flight scenarios.
In the \ac{tl} scenario, while the \ac{vms} detector performs similarly to \ac{zf} detector when $\eta=1$, the \ac{zf} detector becomes superior as soon as $\eta<1$.
In the \ac{cd} phase, the \ac{vms} detector outperforms \ac{zf} when $\eta \geq 0.98$ for $\tau=9$, $\eta \geq 0.99$ for $\tau=17$, and $\eta=1$ for $\tau=25$. \ac{vms} is highly sensitive to \ac{cfo} pre-compensation inaccuracies, with even a small decrease in $\eta$ causing significant performance degradation, whereas \ac{zf} remains more robust under imperfect \ac{cfo}.
Lastly, in the \ac{ec} scenario, the \ac{vms} detector can outperform the \ac{zf} detector at lower values of $\eta$ than in the \ac{cd} scenario. Moreover, as  $\eta$  decreases, the performance of \ac{vms} degrades less dramatically than in the \ac{cd} scenario.


When perfect channel knowledge at the \ac{rx} is assumed, the \ac{vms} detector is expected to outperform the \ac{zf} detector, particularly because the \ac{vms} detector is more resilient to low \ac{snr} conditions, which are typical in \ac{ag} communications \cite{mumimoBook}. 
Comparing the performance of these detectors in Fig.~\ref{fig:accuracy} under imperfect channel knowledge, however, reveals that the advantage of the \ac{vms} detector is strictly tied to the accuracy of the \ac{cfo} pre-compensation. Specifically, the \ac{zf} detector proves more robust and achieves superior performance in scenarios where the \ac{cfo} pre-compensation is inaccurate.
Moreover, while employing larger $\tau$ values is generally expected to improve channel estimation accuracy in low \ac{snr} conditions, the presence of \ac{cfo} significantly complicates this approach.
These findings underscore the necessity of developing a \ac{cfo}-robust channel estimation method, which would enable the use of the \ac{vms} detector with larger $\tau$ values to achieve lower $\mathcal{P}_{\text{out}}$ in low \ac{snr} conditions.

\subsection{Channel Aging Analysis}
We examine channel aging  in Fig.~\ref{fig:time}.  Considering that the estimated channel gradually becomes outdated due to the mobility of the \ac{ac}, this analysis provides insight into how frequently the channel needs to be estimated in a \ac{noma} system for \ac{atm} communications. To specifically focus on the impact of channel aging, we assume perfect \ac{tx} \ac{cfo} pre-compensation accuracy, i.e., $\eta=1$, in these evaluations. We use \ac{zc} sequences of length $\tau=K+1$.

In Fig.~\ref{fig:time}, we observe that, for both the \ac{zf} and \ac{vms} detectors, during the \ac{ec} phase, where the Doppler spread is minimal, $\mathcal{P}_{\text{out}}$ increases very slowly. In this scenario, we see that $\mathcal{P}_{\text{out}}$ remains nearly constant for at least \qty{600}{\milli\second} for $K \leq 8$.
Moreover, the \ac{vms} detector outperforms the \ac{zf} detector in the \ac{ec} scenario.
In contrast, in the \ac{tl} scenario, for both the \ac{zf} and \ac{vms} detectors, $\mathcal{P}_{\text{out}}$ increases sharply over time due to the higher Doppler spread and faster variations in the \ac{aoa} of the \ac{mpc}s. Specifically, we observe that the $\mathcal{P}_{\text{out}}$ of the \ac{vms} detector increases more rapidly than that of the \ac{zf} detector, such that the \ac{zf} detector outperforms the \ac{vms} detector approximately \qty{10}{\milli\second} after channel estimation. We attribute this to the accumulation of residual noise in the \ac{vms} detector.
Moreover, in the \ac{cd} scenario, we observe a sharp increase in $\mathcal{P}_{\text{out}}$ between \qty{3}{\milli\second} and \qty{30}{\milli\second} after channel estimation; however, after \qty{30}{\milli\second}, $\mathcal{P}_{\text{out}}$ increases relatively slowly. In this scenario, the \ac{vms} detector outperforms the \ac{zf} detector initially; but their performance becomes similar approximately \qty{500}{\milli\second} after channel estimation. We also note that the \ac{cd} scenario achieves the lowest $\mathcal{P}_{\text{out}}$ values. This is likely because, while it has lower free-space path loss leading to higher \ac{snr} values compared to the \ac{ec} scenario, it also exhibits considerably lower Doppler spread than the \ac{tl} scenario.

In the \ac{cd} and \ac{ec} phases, $\mathcal{P}_{\text{out}}$ remains low for an extended period after channel estimation. This means that the estimated channel matrix can be used for a long time, and as a result, frequent channel estimation is not required. In turn, this allows a greater portion of available resources to be allocated to actual data transmission, improving overall system efficiency.
Conversely, in the \ac{tl} phase, frequent channel estimation is required, which raises the question of whether the application of a \ac{noma} system is justifiable in this phase.

\section{Conclusion and Outlook}\label{sec:conc}
This paper investigates channel estimation and the outage probability under imperfect channel knowledge for multiple antenna \ac{noma} systems in \ac{atm} communications. 
To accurately model the propagation characteristics, we employ a realistic geometry-based stochastic \ac{ag} channel model, derived from dedicated flight measurement campaigns.
We assume that multiple \ac{ac} transmit simultaneously to the \ac{gs}. 
Our evaluations focuses on two key aspects: 1) channel estimation with imperfect \ac{cfo} pre-compensation at the \ac{tx}, and 2) channel aging, i.e., how the channel evolves over time.
We estimate the channel using \ac{zc} sequences, and based on this estimate, we compute two detectors: 1) \ac{zf} and 2) \ac{vms}. 

The results demonstrate that although the \ac{vms} detector is more robust to low \ac{snr} conditions than the \ac{zf} detector, inaccuracies in \ac{cfo} pre-compensation during channel estimation significantly impair the performance of the \ac{vms} detector, causing it to perform worse than the \ac{zf} detector in certain scenarios. 
Consequently, a \ac{cfo}-robust channel estimation technique is required. Without it, the \ac{vms} detector cannot fully exploit its potential to achieve high spectral efficiency under low \ac{snr} conditions.
Furthermore, the channel aging analysis indicates that while the channel ages rapidly in the \ac{tl} scenario, the channel estimates remain valid over an extended period in the \ac{cd} and \ac{ec} scenarios. These findings suggest that when \ac{cfo}-robust channel estimation is achieved, the \ac{cd} and \ac{ec} phases provide favorable propagation characteristics for implementing a \ac{noma} system.
In our future work, we will focus on further investigating the development of a \ac{cfo}-robust channel estimation technique for \ac{noma}-based \ac{ag} communications.

\bibliographystyle{IEEEtran}

\bibliography{IEEEabrv,ref}

\end{document}